\documentstyle[aps,multicol]{revtex}

\begin{document}

\title{Microscopic Hamiltonian for Zn or Ni substituted high temperature cuprate
superconductors}

\author{T. Xiang$^1$, Y. H. Su$^1$, C. Panagopoulos$^2$, Z. B. Su$^1$ and L.
Yu$^{1,3}$}

\address{$^1$Institute of Theoretical Physics and Interdisciplinary
Center of Theoretical Studies, Academia Sinica, P.O.Box 2735,
Beijing 100080, The People's Republic of China}

\address{$^2$Cavendish Laboratory and IRC in Superconductivity,
University of Cambridge,
Cambridge CB3 0HE, United Kingdom}

\address{$^3$ Abdus Salam International Centre for Theoretical Physics,
P.O.Box 586, 34100 Trieste, Italy}

\maketitle

\begin{abstract}

We have derived the effective low energy Hamiltonian for Zn or Ni
substituted high-T$_c$ cuprates from microscopic three-band models
consisting of the most relevant Cu or impurity $3d$ and O $2p$
orbitals. We find that both scattering potential and hopping
integral induced by impurities have a finite range but decay very
fast with distance from the impurity. The Zn scattering potential
is very strong and attractive for electrons. The Ni scattering
potential is much weaker than the Zn case, resulting from the
hybridization between Ni ions and O holes. This profound
difference is due to neither the  electric charge nor $d$-level
location, but rather because of the interplay between the valence
state of the impurity and the strong correlation background. It
gives a natural account for the unusual effect of Ni and Zn on the
reduction of superconducting transition temperature. The
interlayer hopping of electrons is highly anisotropic and
nonlocal, determined by the in-plane electronic structure. This
leads to a quantum interference of states from different sites and
affects strongly the scanning tunneling spectrum perpendicular to
CuO$_2$ planes.

\end{abstract}

\begin{multicols}{2}

Substitution of divalent transition metal Zn or Ni for Cu in Cu-O plane
 has offered a fruitful and challenging direction in exploring the
nature of correlated electrons in high-T$_c$ cuprates\cite
{Xia87,Chi91,Bon94,Mah94,Pan00,Rad93,Bor94,Xia95,Bal95,Zha01,Pol01,Mar02,Zhu00,Poi94,Fin90,Nag95}.
The effects of these impurities on superconducting properties are
unusual and in many ways opposite to those observed in
conventional superconductors. In particular, the T$_c$ reduction
induced by non-magnetic Zn ions is almost three times larger than
that by magnetic Ni ions\cite{Chi91}. Also, the temperature
dependence of the penetration depth shows a much stronger
pair-breaking effect of Zn compared with Ni\cite{Bon94}. The
actual situation is more complicated than it appears: the NMR and
susceptibility measurements  do show the existence of an induced
local moment around Zn impurities\cite{Mah94}.
 There exists a vast literature devoted to explore the impurity effects on
$d$-wave
superconductors\cite{Rad93,Bor94,Xia95,Bal95,Ope98,Kap02,Lia02},
particularly to interpret the scanning tunneling microscopy (STM)
experiments\cite{Pan00,Zha01,Pol01,Mar02,Zhu00}.  However, the key
issue, namely why Zn impurity is a very strong scatterer (close to
the unitary limit), while Ni is a relatively weaker scatterer, is
not well understood. The earlier efforts  to address this issue
using exact diagonalization of the $t-J$ model on finite
clusters\cite{Poi94} and the slave-particle treatments of the
resonance valence bond (RVB) state\cite{Fin90,Nag95} show the
importance of the strong correlation background, but are unable to
elucidate the physics involved  to the full extent.

In this paper, we present a theoretical derivation for the
effective low-energy Hamiltonians for both Zn and Ni impurity
systems. The aim is to find an accurate description for the
impurity scattering potential and to clarify a number of important
issues associated with the scattering phase shifts. We also
present a detailed analysis for the anisotropic interlayer hopping
process and discuss its impact on the scanning tunneling spectra
of high-T$_c$ superconductors. We start from the three-band
Hubbard model for the Cu-O plane, and study systematically the
influence of the impurity on the Zhang-Rice singlet\cite{Zha88}
formation. In this work, only a single Zn or Ni impurity system is
studied. However, it is straightforward to extend the derivation
to a system of many Zn or Ni impurities, provided  the direct
interaction among them is negligible.

Let us first consider the case of Zn impurity. Zn$^{2+}$ has a
3d$^{10}$ configuration and no spin. Since the valence fluctuation
of Zn$^{2+}$ is small, we can model Zn as an inert non-magnetic
impurity with no coupling with surrounding O or Cu states. In
high-T$_c$ oxides, the low energy physics is governed by one Cu
$3d_{x^2-y^2}$ and two O ($2p_x,2p_y$) bands and the doped holes
are predominantly on O ($2p_x,2p_y$ ) states. Thus to study the
physical effect of Zn, we should start from the following
three-band Hamiltonian
\begin{eqnarray}
H &=&H_0+H_1,  \label{ham1} \\
H_0 &=&\varepsilon _p\sum_lp_l^{\dagger }p_l+\sum_{i\neq
i_0}\left(\varepsilon_dd_i^{\dagger }d_i+U_dd_{i\uparrow }^{\dagger
}d_{i\uparrow }d_{i\downarrow }^{\dagger }d_{i\downarrow }\right),
\nonumber \\
H_1 &=&-\sum_{\left\langle il\right\rangle i\neq i_0}t_{pd}\left(
p_l^{\dagger }{}d_i+d_i^{\dagger }p_l\right) ,  \nonumber
\end{eqnarray}
where $d_i^{\dagger }=\left( d_{i\uparrow }^{\dagger },d_{i\downarrow
}^{\dagger }\right) $ creates a Cu $3d_{x^2-y^2}$ hole, and $p_l^{\dagger
}=\left( p_{l\uparrow }^{\dagger },p_{l\downarrow }^{\dagger }\right) $
creates an O $2p_x$ hole if $l=i\pm \widehat{x}/2$ or an O $2p_y$ hole if
$l=i\pm \widehat{y}/2$. (See Fig. 1 of Ref. \cite{Xia98} for the
definition of relative phases of $3d_{x^2-y^2}$ and O $2p$ orbitals.)
 $i_0$ is the   impurity position. Without loss of
generality, we set $\varepsilon _d=0$ and $\varepsilon _p>0$. The
real system is in the limit $U_d\gg \varepsilon _p\gg t_{pd}$. An on-site
Coulomb repulsion term for O holes is neglected since we are only interested
in low doping cases and the probability for two holes to occupy one O site
is very small.

The O $2p$ holes can be classified according to their hybridization with Cu
$3d_{x^2-y^2}$ states as bonding and non-bonding Wannier orbitals. These
orbitals are defined on the Cu lattice. The bonding orbital is defined by
$a_i=\frac 1{\sqrt{N}}\sum_ke^{ikR_i}\beta _k^{-1}\left[ \cos
(k_x/2)p_{x,k}+\cos (k_y/2)p_{y,k}\right] $, where $\beta _k=\sqrt{\cos
^2(k_x/2)+\cos ^2(k_y/2)}$ and $N$ is the number of Cu sites. $p_{x,k}$ and
$p_{y,k}$ are the Fourier transformation of $p_{i+\widehat{x}/2}$ and $p_{i+
\widehat{y}/2}$, respectively. The non-bonding orbital is orthogonal to $a_i$
and defined by $b_i=\frac 1{\sqrt{N}}\sum_ke^{ikR_i}\beta _k^{-1}\left[ \cos
(k_y/2)p_{x,k}-\cos (k_x/2)p_{y,k}\right] $. With these Wannier orbitals,
the Hamiltonian (\ref{ham1}) can be expressed as
\begin{eqnarray*}
H_0 &=&\varepsilon _p\sum_i\left( a_i^{\dagger }a_i+b_i^{\dagger }b_i\right)
+\sum_{i\neq i_0}U_dd_{i\uparrow }^{\dagger }d_{i\uparrow }d_{i\downarrow
}^{\dagger }d_{i\downarrow }, \\
H_1 &=&-t_{pd}\sum_{i\neq i_0,j}u\left( i-j\right) \left( a_j^{\dagger
}d_i+d_i^{\dagger }a_j\right) ,
\end{eqnarray*}
where $u\left( i\right) =\frac 2N\sum_k\beta _ke^{ikR_i}$. The non-bonding
orbitals do not couple to the other states and therefore may be neglected.

In the limit $U_d\gg \varepsilon _p\gg t_{pd}$, we can treat $H_1$ as a
perturbation and use degenerate perturbation theory to project the
Hamiltonian onto the Hilbert space spanned by the low-energy states of $H_0$. Up
to the second order of perturbation, we find that an effective Hamiltonian
for the system is given by
\begin{equation}
H_{Zn}^{^{\prime }}=-\sum_{ij}t_{ij}a_i^{\dagger }a_j+t_{12}\sum_{i\neq
i_0\sigma \sigma ^{\prime }}d_{i\sigma }^{\dagger }d_{i\sigma ^{\prime }}
\widetilde{a}_{i\sigma ^{\prime }}^{\dagger }\widetilde{a}_{i\sigma },
\end{equation}
where $t_{12}=t_1+t_2$, $t_1=t_{pd}^2/\varepsilon _p$, $t_2=t_{pd}^2/(U_d-\varepsilon _p)$,
$t_{ij}=t_2\left[ 4\delta _{i,j}+\delta _{\left\langle
i,j\right\rangle }-u(i_0-i)u(i_0-j)\right] $ (the second Kronecker symbol
 means $i,j$ must  be the nearest neighbours) and
$\widetilde{a}_i=\sum_ju\left( i-j\right) a_j$.
The coupling between $d_i$ and $a_j$ at
different sites is much smaller than the on-site one. Thus, we can take the
approximation $\widetilde{a}_i\simeq u\left( 0\right) a_i$. This is also an
approximation implicitly assumed in the work of Zhang and Rice\cite{Zha88}.

In $H^{\prime }$, the coupling constant between O $2p$ and Cu $3d_{x^2-y^2}$
states is of the order $\left( t_2+t_1\right) u^2(0)$. This is a large energy
scale compared with the hopping integral for the bonding O holes. An
important consequence of this strong coupling between O $2p$ and Cu
$3d_{x^2-y^2}$ holes, as first pointed out by Zhang and Rice\cite{Zha88}, is
that an O hole will form a local spin singlet with a Cu spin. The binding
energy of the Zhang-Rice singlet is $\left( t_2+t_1\right) u^2(0)$. The
energy difference between the Zhang-Rice singlet and the corresponding
triplet is $2\left( t_2+t_1\right) u^2(0)$. Since this difference is much
larger than the bandwidth of the O holes, the triplet hole band can be neglected
in the study of low energy excitations. Thus we can further project the
Hamiltonian onto the subspace spanned by the Zhang-Rice singlets with
unpaired Cu spins. The resulting effective Hamiltonian is then given by
\begin{equation}
H_{Zn}^{\prime \prime }=\sum_iV_{Zn}(i)d_i^{\dagger }d_i-\sum_{i\neq
j}t_{ij}^{Zn}d_j^{\dagger }d_i  \label{ham2}
\end{equation}
subjected to the constraint $d_i^{\dagger }d_i\leq 1$ when $i\neq i_0$. At
the impurity site, $d_{i_0}=\left( d_{i_0\uparrow },d_{i_0\downarrow
}\right) $ is defined from the creation operator of the bonding O hole:
$d_{i_0\sigma }=-\sigma a_{i_0\overline{\sigma }}^{\dagger }$. We should
emphasize that $d_{i_0}$ is not an annihilation operator for the Zn
$3d_{x^2-y^2}$ state. At the Zn site, there is no constraint imposed on
$d_{i_0}$. Thus two electrons with opposite spins can occupy this site
simultaneously. This is consistent with the fact that there are two
electrons in the $3d_{x^2-y^2}$ state of the Zn$^{2+}$ ion, although our
starting Hamiltonian ( \ref{ham1}) does not contain that state at all.
Moreover, if the total number of doped holes in the system is $N_h$, then it
can be shown that $\sum_id_i^{\dagger }d_i=(N-N_h)+1$. Thus the total number
of conducting electrons is more than the nominal charge of the system by
one. This implies that Zn$^{2+}$ can be {\em effectively} taken as an ion
with one more charge than Cu$^{2+}$, although both are divalent. It further
suggests that we only need to count the difference between the number of 3d
electrons in Zn$^{2+}$ and Cu$^{2+}$ when the Friedel sum rule of phase
shifts is used.

In (\ref{ham2}), the hopping integral is defined by
\begin{equation}
t_{ij}^{Zn}=\frac{\widetilde{t}_{ij}}2\delta _{i\neq i_0,j\neq i_0}+\frac{
\widetilde{t}_{i_0j}}{\sqrt{2}}\delta _{i,i_0}+\frac{\widetilde{t}_{i_0i}}{
\sqrt{2}}\delta _{j,i_0},
\end{equation}
where $\widetilde{t}_{ij}=t_2\delta _{\left\langle i,j\right\rangle
}-t_2u(i_0-i)u(i_0-j)$. The first term in $\widetilde{t}_{ij}$ is the usual
hopping integral in a homogeneous system, and the second term is a small
correction to the hopping integral by the Zn impurity. This impurity induced
hopping integral is non-local but decays very fast with distance from the
impurity.

The impurity potential $V_{Zn}(i)$ is given by
\begin{equation}
V_{Zn}(i)=-t_2u^2(i_0-i)\delta _{i\neq i_0}-\left( 2t_2+t_1\right) u^2\left(
0\right) \delta _{i,i_0}.
\end{equation}
This is an attractive potential. It is certainly not of a
$\delta$-function form, but away from the Zn impurity, $V_{Zn}(i)$
decays very fast, roughly as $|i-i_0|^{-6}$. At the impurity site,
$|V_{Zn}(i_0)|\approx 3.67\left( t_1+2t_2\right) $ is about two
orders of magnitude larger than the potential at the four nearest
neighboring sites of Zn, $|V_{Zn}(i)|\approx 0.0785t_2$. It is
also more than one order of magnitude larger than the hopping
constant $t_2/2$, in consistent with the STM measurement
data\cite{Pan00,Mar02}. This shows clearly that Zn impurity is
indeed a strong potential scatterer.

The strong attractive potential induced by Zn is in fact due to a strong
repulsion of Zn$^{2+}$ to the bonding O holes at the impurity site. This
strong repulsion of Zn$^{2+}$ to O holes results from two distinguished
physical effects. First, since Zn$^{2+}$ has no spin, O holes cannot gain
energy by forming a Zhang-Rice singlet with a $3d_{x^2-y^2}$ spin at the Zn
site, this costs the O hole an energy of $(t_2+t_1)u^2\left( 0\right) $.
Second, since there is no hopping between O 2p and Zn 3d states, the O hole
also suffers from a kinetic energy loss of $t_2u^2\left( 0\right) $. The
total energy loss of an O hole at the impurity site is thus $(2t_2+t_1)u^2\left(
0\right) $. This serves effectively as a repulsive potential to the O holes,
or equivalently an attractive potential to the $d_{i_0}$ electrons, at the
Zn site. This picture is completely different from the naive interpretation
for the origin of the attractive potential of Zn in literature, which claims
that the potential is attractive because Zn$^{2+}$ has one more electron in
its $3d_{x^2-y^2}$ orbital than Cu$^{2+}$ and the $3d$ energy levels of
Zn$^{2+}$ are lower than those of Cu$^{2+}$. We would like to
particularly point out  an intriguing connection
between the scattering potential of Zn and the physics underlying the
Zhang-Rice singlet. It suggests that through the measurement of the Zn
scattering potential, we can even gain direct information on the binding
energy of the Zhang-Rice singlet.

The Hamiltonian (\ref{ham1}) contains the most important information about
the Zn scattering potential. However, to obtain the full effective
Hamiltonian, we should include the fourth order of perturbation as in the
derivation of the standard $t-J$ model. In particular, two terms which are
absent in the second order of perturbation should be included. One is the
exchange energy between two Cu$^{2+}$ spins on nearest neighboring sites.
The other is the hopping of electrons between next or next-next nearest
neighboring sites. This term has strong effect on the structure of Fermi
surface. There is also a three-site hopping term, which is generally ignored
for simplicity. The other terms generated from the fourth order of
perturbation are small corrections to the terms in (\ref{ham2}) and can be
omitted. Thus the full effective Hamiltonian for the Zn-substituted system is

\begin{equation}
H_{Zn}=H_{Zn}^{\prime \prime }-\sum_{ij\neq i_0}t_{ij}^{\prime }d_i^{\dagger
}d_j+\sum_{\left\langle ij\right\rangle \neq i_0}JS_i\cdot S_j,  \label{zn}
\end{equation}
where $S_i$ is the Cu spin operator, and $t_{ij}^{\prime
}=t^{\prime }$ if $(i,j)$ are next neighbors, or $t^{\prime \prime
}$ if $(i,j)$ are next-next neighbors, or zero otherwise.
$t^{\prime } $ and $t^{\prime \prime }$ always have opposite
signs. They are of the same order as the exchange energy $J$.

Now let us turn to the Ni case. As for $Zn^{2+}$, $Ni^{2+}$ is also in a
divalent state. It has 8 electrons in its $3d$ states. In particular, it has
one electron in the $3d_{3z^2-r^2}$ state and one in the $3d_{x^2-y^2}$
state. These two electrons form a spin triplet via a strong Hund's rule
coupling. Like $Zn^{2+}$, $Ni^{2+}$ is also very stable. However, unlike
$Zn^{2+}$, the hybridization between the Ni $\left(
3d_{x^2-y^2},3d_{3z^2-r^2}\right) $ and O ($2p_x,2p_y$) states is strong
since there is only one electron in each of these $3d$ states.

The appropriate three-band model for the Ni system is defined by the
Hamiltonian
\begin{eqnarray}
H_{Ni} &=&H+\sum_\alpha \varepsilon _\alpha ^{Ni}c_\alpha ^{\dagger
}c_\alpha -\sum_{\left\langle li_0\right\rangle \alpha }t_\alpha ^{Ni}\left(
p_l^{\dagger }{}c_\alpha +h.c.\right)  \nonumber \\
&&-J_Hc_1^{\dagger }\frac \sigma 2c_1\cdot c_2^{\dagger }\frac \sigma
2c_2+\sum_\alpha U_\alpha ^{Ni}c_{\alpha \uparrow }^{\dagger }c_{\alpha
\uparrow }c_{\alpha \downarrow }^{\dagger }c_{\alpha \downarrow },
\label{ni1}
\end{eqnarray}
where $H$ is the same as defined in (\ref{ham1}). $\alpha =1$ or $2$ denotes
the $Ni^{2+}$ $3d_{x^2-y^2}$ or $3d_{3z^2-r^2}$ state. $c_\alpha =(c_{\alpha
\uparrow },c_{\alpha \downarrow })$ is the electron operator for these
$Ni^{2+}$ states. $\varepsilon _\alpha ^{Ni}$, $U_\alpha ^{Ni}$, and
$t_\alpha ^{Ni}$ are the energy level, the on-site Coulomb repulsion, and the
hopping integral of the corresponding $3d$ state of $Ni^{2+}$, respectively.
$J_H$ is the Hund's rule coupling constant.

In (\ref{ni1}), $t_\alpha ^{Ni}$ is much smaller than $J_H$. Thus this
$t_\alpha ^{Ni}$ term can be also treated as a perturbation. The effective
one-band Hamiltonian for the Ni impurity can be derived following the same
procedure as for the Zn impurity. The difference is that an O hole can now
form a local spin doublet (an analogue to the Zhang-Rice singlet) with the
Ni$^{2+}$ spin at the impurity site. This difference is quite important, and it
leads to a relatively weaker scattering potential for Ni. The effective
one-band Hamiltonian we obtained for the Ni impurity is given by
\begin{eqnarray}
H_{Ni} &=&\sum_{i\neq j\neq i_0}\frac{t_{ij}^{Ni}}2d_j^{\dagger
}d_i-\sum_{i\neq i_0}\frac{t_{i_0i}^{Ni}}{\sqrt{3}}\left( T_i^{\dagger
}T+h.c.\right)  \nonumber \\
&&+\sum_iV_{Ni}(i)d_i^{\dagger }d_i+\sum_{\left\langle ij\right\rangle,
}J_{ij}S_i\cdot S_j  \label{ni2}
\end{eqnarray}
where $T=\left( T_1,T_0,T_{-1}\right) $ is the annihilation
operator for the three spin triplet states of Ni$^{2+}$, and
$d_{i_0}=(d_{i_0\uparrow },d_{i_0\downarrow })$ is the
annihilation operator of the local spin doublet formed by the
Ni$^{2+}$ spin and the bonding O hole at the impurity site. As for
the Zn case, $d_{i_0}$ is not an electron operator for the $3d$
state of Ni$^{2+}$. $T_i^{\dagger }\equiv \left( d_{i_0\uparrow
}^{\dagger }d_{i\uparrow }^{\dagger },\left( d_{i_0\uparrow
}^{\dagger }d_{i\downarrow }^{\dagger }+d_{i_0\downarrow
}^{\dagger }d_{i\uparrow }^{\dagger }\right) /
\sqrt{2},d_{i_0\downarrow }^{\dagger }d_{i\downarrow }^{\dagger
}\right) $. When $i\neq i_0$, the constraint $d_i^{\dagger
}d_i\leq 1$ is imposed. At the impurity site, the constraint
becomes $d_{i_0}^{\dagger }d_{i_0}+T^{\dagger }T=1$, namely the
impurity site must be either in a spin triplet state of pure
Ni$^{2+}$ or in a Zhang-Rice-type spin doublet state. The second
term in $H_{Ni}$ describes a swap process between the spin triplet
and the spin doublet at the Ni site by absorbing or releasing an
electron at the other site. The exchange constant $J_{ij}=J$ if
both $i$ and $j$ are not equal to $i_0$. For the bonds connecting
the Ni site $J_{ij}$ is different from but of the same energy
scale as J.

The hopping integral $t_{ij}^{Ni}$ is
\begin{equation}
t_{ij}^{Ni}=t_{ij}^0+\left( t_4-t_2\right) u(i_0-i)u(i_0-j),
\end{equation}
where $t_{ij}^0$ is the hopping integral as in the standard $t-J$ model
without impurities, including the hopping between nearest, next nearest and
next-next nearest neighbors. The second term in $t_{ij}^{Ni}$ is the hopping
integral induced by the Ni impurity. This term is smaller than the
corresponding term in the Zn case.

The scattering potential of Ni reads
\begin{eqnarray}
V_{Ni}(i) &=&\left( 2t_2+t_1-\frac 12t_3-\frac 32t_4\right) u^2(0)\delta
_{i,i_0}  \nonumber \\
&&-\left( t_2-t_4\right) u^2(i_0-i)\delta _{i\neq i_0},
\end{eqnarray}
where $t_3=\sum_\alpha \left( t_\alpha ^{Ni}\right) ^2/(\varepsilon
_p-\varepsilon _\alpha ^{Ni}+J_H/4)$ and $t_4=\sum_\alpha \left( t_\alpha
^{Ni}\right) ^2/(U_\alpha ^{Ni}-\varepsilon _p+\varepsilon _\alpha
^{Ni}+J_H/4)$. The $t_3$ term originates from the hopping between $Ni^{2+}$
$3d$ and O $2p$ states. The $t_4$ term comes from the energy gain by the
formation of the local spin doublet at the impurity site. These terms
compensate or even overcome the energy loss resulted from the $t_1$ and
$t_2 $ terms. Thus the scattering potential of Ni is much weaker than Zn.

The total number of unpaired electrons is now determined by the formula
$\sum_{i\neq i_0}d_i^{\dagger }d_i=N-N_{hole}-T^{\dagger }T$. If the
scattering potential is strong and repulsive to the spin doublet at the Ni
site, then this site is essentially in the spin triplet state, i.e.
$T^{\dagger }T=1$, and the conducting electron number is one less than the
total number of charge. In this case the Friedel sum rule should be used as
if one charge is absorbed by the Ni$^{2+}$ ion. If, on the other hand,
scattering potential is strong but attractive to the spin doublet at the Ni
site, then this site is predominantly in the spin doublet state and the
total number of conducting electrons is equal to the total number of charge.
In this case, the scattering phase shifts of high angular momentum channels
become important and the commonly adopted T-matrix approximation\cite{Hir88}
is no longer applicable. In real systems, the Ni site might be in a mixed valence
state, with finite probabilities in both spin triplet and doublet states. In
this case, a thorough analysis of the Ni scattering becomes very
complicated, even without considering its magnetic coupling with Cu spins.

By comparing the results for Zn and Ni cases we clearly see the
difference: Zn is a very strong scatterer, while Ni is a weak one.
This profound difference is
 due to neither the  electric charge nor $d$-level location, but rather because of the
interplay between the valence state of the impurity and the strong
correlation background. The underlying physics is best revealed
using the three band model and the concept of Zhang-Rice singlet.
The penetration depth measurements give us strong hints that Zn
materializes a unitary scattering, whereas Ni is a scatterer close
to the Born scattering limit\cite{Bon94}. The present microscopic
derivation provides a very good justification for this scenario.
We did not calculate explicitly the magnetic moment distribution
and the contribution of spin-flip scattering around a Zn or Ni
impurity. However, we believe that these problems can be addressed
by self-consistently solving the effective Hamiltonians for Zn or
Ni impurities . The fact that the suppression of T$_c$ scales
universally with the increase of the residual resistance
\cite{Chi91}, and the rough estimate of the dominance of the
potential scattering over magnetic scattering\cite{Bor94}, do
support that the spin-flip scattering induced by Zn or Ni
impurities is a secondary effect compared with  the potential scattering,
especially for the Zn case.

Finally, let us discuss the problem about the interlayer hopping
of electron around an impurity. This problem is of great
importance for the understanding of various experimental results
related to impurities, especially the scanning tunneling data in
the superconducting state around Zn or Ni\cite{Pan00}. In
high-T$_c$ materials, the interlayer hopping is fulfilled through
a virtual hopping process assisted by Cu $4s$ orbitals. That is an
O hole first hops to a Cu $4s$ orbital and then hops to other
orbitals out of the CuO$_2$ plane\cite{Xia98,Xia96}. At the
impurity site, the interlayer hopping is assisted by the Zn $4s$
or Ni $4s$ state. Since the $4s$ orbital is orthogonal to the
$3d_{x^2-y^2}$ orbital at the same site, there is no hopping
between these two orbitals. This $4s$-orbital assisted hopping has
profound consequences on c-axis dynamic properties\cite
{Xia96,Xia98}.

The $4s$-orbital assisted hopping in each CuO$_2$ layer is
described by the following Hamiltonian:
\begin{equation}
H_c=\sum_{\left\langle il\right\rangle }\left( -\right) ^{\alpha
_{li}}t_i^{4s}\left( \widetilde{s}_i^{\dagger }p_l+p_l^{\dagger
}{} \widetilde{s}_i\right)
\end{equation}
where $\left( -\right) ^{\alpha _{li}}=1$ if $l=i\pm
\widehat{x}/2$ and $ \left( -\right) ^{\alpha _{li}}=-1$ if
$l=i\pm \widehat{y}/2$. $\widetilde{s} _i^{\dagger
}=(\widetilde{s}_{i\uparrow }^{\dagger },
\widetilde{s}_{i\downarrow }^{\dagger })$ creates a hole in the
local $4s$ state. $t_i^{4s}$ is the hopping integral between a
$4s$ and its neighboring O $2p$ orbitals. By projecting this
Hamiltonian onto the low energy subspace containing Zhang-Rice
singlets, we find that $H_c$ can be expressed as
\begin{equation}
H_c=\sum_it_i^{4s}s_i^{\dagger }D_i+h.c.  \label{caxis}
\end{equation}
where $s_i^{\dagger }=(-\widetilde{s}_{i\downarrow },\widetilde{s}
_{i\uparrow })$ is the electron creation operator for the $4s$
state at site $i$,
\begin{equation}
D_i=\sum_jF_{i,j}\left( \delta _{j,i_0}+\frac 1{\sqrt{2}}\delta
_{j\neq i_0}\right) d_j  \label{d1}
\end{equation}
for the Zn impurity model, and
\begin{equation}
D_i=\sum_{j\neq
i_0}\frac{F_{i,j}}{\sqrt{2}}d_j-\frac{F_{i,i_0}}{\sqrt{3}}
d_{i_0}^{\dagger }\left(
\begin{array}{cc}
\sqrt{2}T_1, & T_0 \\
T_0, & \sqrt{2}T_{-1}  \end{array} \right)  \label{d2}
\end{equation}
for the Ni impurity model. In Eqs. (\ref{d1}) and (\ref{d2}),
\begin{equation}
F_{i,j}=\frac 1N\sum_k\beta _k^{-1}(\cos k_x-\cos k_y)e^{ik(i-j)}.
\end{equation}
It is straightforward to show that $F_{i,j}=0$ when $i=j$ or when
$i-j$ is along two diagonal directions. Thus in the effective
one-band model, an electron in the CuO$_2$ plane cannot hop to a
$4s$ orbital at the same site. This is a peculiar but important
feature of the c-axis hopping integral.

The local tunneling current perpendicular to CuO$_2$ planes as
measured by the STM is proportional to the probability of
electrons in the CuO$_2$ plane hopping to the local $4s$ orbital
from the low energy conducting band. From Eq. (\ref{caxis}) one
can show that the tunneling conductance at site $i$ is determined
by the following non-local matrix element
\begin{equation}
g_i(V)\propto \left\langle D_i\delta (eV-H_{imp})D_i^{\dagger
}\right\rangle \label{con}
\end{equation}
where $V$ is the applied bias voltage and $H_{imp}$ is the system
Hamiltonian as given in Eqs. (\ref{zn}) or (\ref{ni2}). This
formula shows clearly that the scanning tunneling conductance is
not simply a measure of the local density of states in high-T$_c$
superconductors. In particular, as $F_{i,j}=0$ when $i=j$, there
is no contribution to $g_i(V)$ from the low energy excitations at
site $i$. This is a peculiar but intrinsic property of the
in-plane electronic structure. It is not due to the blocking
effect as suggested in Ref. \cite{Zhu00}. In Eq. (\ref{d1}) or
(\ref{d2}), if $j$ in the summation only takes the four nearest
neighboring sites of $i$, we then obtain the result given by
Martin et al\cite{Mar02}. Their result is a good approximation to
Eq. (\ref{con}) since $F_{i,j}$ decays very fast with the distance
$\left| j-i\right| $. Eq. (\ref {con}) is valid independent on the
structure of the insulating layers between the top CuO$_2$ plane
and the STM tip. Thus the normalized STM spectrum should be the
same no matter whether the tip is above an insulating layer or
directly above a CuO$_2$ layer.

In conclusion, we have derived the effective low energy
Hamiltonian with the impurity scattering potential for Zn or Ni
substituted high-T$_c$ cuprates. Our results reveal an intrinsic
connection of the impurity scattering potential with the
correlation effect of electrons and set a solid microscopic
starting point towards thorough understanding of the impurity
effects in these materials. A number of observable physical
quantities can be calculated explicitly based on this microscopic
model. The STM spectrum perpendicular to CuO$_2$ is not simply a
probe of the local density of states since the interlayer hopping
is intrinsically anisotropic and non-local.

We thank G. M. Zhang for useful discussions. TX acknowledges the
hospitality of the Interdisciplinary Research Center in
Superconductivity of the University of Cambridge, where part of
the work was done, and the financial support from the National
Natural Science Foundation of China and from the special funds for
Major State Basic Research Projects of China.



\end{multicols}

\end{document}